\documentclass[10pt]{article} % For LaTeX2e

\usepackage[accepted]{rlj} % Camera-ready / accepted version
\usepackage{amssymb}            % Defines common symbols like \mathbb R
\usepackage{mathtools}          % Extends amsmath, providing common math tools
\usepackage{booktabs}           % For \toprule, \midrule, \bottomrule in tables
\usepackage{graphicx}           % For including images
\usepackage{subcaption}         % For subfigure environment
\usepackage{url}                % For displaying URLs
\usepackage{algorithm}          % For algorithm float environment
\usepackage{algorithmic}        % For algorithmic pseudocode commands
\usepackage{cleveref}           % For \cref (must come after hyperref, which rlj loads)

\definecolor{bestblue}{HTML}{2563EB}
\definecolor{impgreen}{HTML}{059669}

\newcommand{\fcTS}{\textsc{FC-TS}}
\newcommand{\swTS}{\textsc{SW-TS}}
\newcommand{\dTS}{\textsc{D-TS}}
\newcommand{\rTS}{\textsc{R-TS}}
\newcommand{\stdTS}{\textsc{Std-TS}}
\newcommand{\best}[1]{\textbf{\textcolor{bestblue}{#1}}}
\newcommand{\imp}[1]{\textcolor{impgreen}{(#1\%)}}
\newcommand{\reals}{\mathbb{R}}
\newcommand{\cH}{\mathcal{H}}
\newcommand{\cN}{\mathcal{N}}

\newcommand{\cI}{\mathcal{I}}
\newcommand{\cR}{\mathcal{R}}

\DeclareMathOperator*{\argmax}{arg\,max}
\DeclareMathOperator*{\argmin}{arg\,min}

\title{Flow-Corrected Thompson Sampling for \\ Non-Stationary Contextual Bandits}

\setrunningtitle{Flow-Corrected Thompson Sampling for Non-Stationary Contextual Bandits}

\author{AmirHossein Naghdi\textsuperscript{1}, Ali Baheri\textsuperscript{2}}

\emails{Amirhosseintt2003@gmail.com, akbeme@rit.edu}

\affiliations{
$^{1}$\textbf{Independent Researcher}\\
$^{2}$\textbf{Rochester Institute of Technology, USA}
}

\contribution{
    We propose Flow-Corrected Thompson Sampling (\fcTS{}), a transport-based extension of Thompson Sampling for non-stationary linear contextual bandits that corrects historical rewards into present-time pseudo-observations and incorporates them via confidence-weighted Bayesian updates rather than uniform forgetting.
    }
    {
    Our method is designed for settings where non-stationarity exhibits exploitable structure (e.g., trend, periodicity, recurring regimes); when drift is unstructured or rapidly adversarial, standard windowing/discounting baselines may be more appropriate.
    }

\contribution{
    We instantiate \fcTS{} for three common drift patterns - linear drift (online ridge regression and linear reward transport), periodic variation (phase-aware reuse across cycles), and recurring regimes (changepoint detection with regime-specific posterior memory) - and provide an adaptive variant that combines trend and multiple periodicities.
    }
    {
    These instantiations rely on model choices (e.g., window length, phase bandwidth, detection thresholds) that must be selected; while our experiments show robustness across reasonable ranges, performance can degrade under severe misspecification.
    }

\contribution{
    We empirically validate the approach on five case studies spanning synthetic and semi-synthetic benchmarks, including trend, periodicity, recurring regimes, compound non-stationarity, and portfolio selection.
    }
    {
    The current study emphasizes algorithmic design and empirical evaluation; broader stress tests under highly misspecified or adversarial drift remain open directions.
    }


\hypersetup{
    pdftitle={Flow-Corrected Thompson Sampling for Non-Stationary Contextual Bandits},
    pdfauthor={AmirHossein Naghdi and Ali Baheri},
    pdfkeywords={contextual bandits, non-stationary bandits, Thompson sampling, distribution shift}
}

\summary{
We study non-stationary linear contextual bandits, where reward parameters drift over time and naive reuse of historical data becomes biased. We propose \emph{Flow-Corrected Thompson Sampling} (\fcTS{}), which transports past rewards to the current time under a drift model and incorporates transported observations with confidence weights. The resulting framework specializes to linear drift, periodic variation, and recurring regime switches, and supports combining trend and seasonality. Across five case studies (including a semi-synthetic portfolio benchmark), \fcTS{} consistently improves cumulative regret over strong forgetting-based baselines such as discounted, sliding-window, and restarting Thompson sampling, with the largest gains when temporal structure recurs.
}

\begin{document}

% \makeCover  % Cover page omitted for workshop submission (optional per RLC format)
\maketitle  % Make the title section

\begin{abstract}
We study non-stationary linear contextual bandits where the reward model drifts over time, rendering classical contextual bandit algorithms brittle because historical data becomes systematically biased. We propose \emph{Flow-Corrected Thompson Sampling} (\fcTS{}), a Bayesian method that reuses experience by \emph{transporting} past rewards to the present using an explicit drift model and incorporating each transported observation with a confidence weight that reflects transport reliability. This yields a unified template that specializes in (i) linear parameter drift via online slope estimation and reward correction, (ii) periodic variation via phase-aware reuse across cycles, and (iii) recurring regime switches via changepoint detection and regime-specific posterior memory. The resulting posterior updates remain closed-form under a linear Gaussian model and can be implemented efficiently with truncated, incrementally updated sufficient statistics. Across five controlled case studies and a semi-synthetic portfolio-selection benchmark with multiple overlapping non-stationarities, \fcTS{} outperforms standard forgetting-based baselines (discounting, sliding windows, and periodic restarts), with the largest gains in settings exhibiting recurring temporal structure. These results demonstrate that when non-stationarity is structured, correcting and reweighting historical observations can be substantially more sample-efficient than uniformly discarding them.
\end{abstract}

\section{Introduction}
\label{sec:intro}

Sequential decision-making under uncertainty lies at the heart of many modern applications, from clinical trials and recommendation systems to online advertising, dynamic pricing, robotics, aviation, and finance \cite{kober2013reinforcement,afsar2022reinforcement,razzaghi2024survey}. The contextual bandit framework provides a principled approach to these problems, enabling learners to balance the exploration of uncertain options with the exploitation of accumulated knowledge \cite{bouneffouf2020survey,baheri2023llms}. A key assumption underlying classical contextual bandit algorithms is stationarity: the relationship between contexts, actions, and rewards remains fixed over time. However, this assumption is routinely violated. User preferences evolve, market conditions shift, and treatment effects change as populations adapt. Ignoring such non-stationarity leads to policies that perform well initially but degrade as the environment drifts away from historical patterns.

The challenge of non-stationary bandits has attracted considerable attention, with existing approaches falling broadly into three categories. Sliding window methods maintain only recent observations, discarding older data that may no longer reflect current conditions \citep{garivier2011upper,cheung2019learning}. Discounted methods assign exponentially decaying weights to past observations, allowing old data to fade gradually from influence \citep{raj2017taming,russac2019weighted}. Restarting methods periodically reset the learning algorithm, treating each epoch as a fresh problem \citep{besbes2014stochastic,auer2019adaptively}. While these approaches successfully adapt to changing environments, they share a fundamental limitation: they treat historical data as a liability to be managed rather than an asset to be leveraged.

We propose a different perspective. When the structure of non-stationarity is predictable, as when preferences drift gradually, follow seasonal patterns, or alternate between recurring regimes, historical observations contain valuable information about the current environment. The key insight is that past rewards can be \emph{transported} to the present by correcting for the intervening drift. An observation that would be misleading if used directly becomes informative once we account for how the environment has changed since it was collected. This transport principle enables our algorithm to maintain a much larger effective sample size than methods that simply discard or downweight old data.

We propose \emph{Flow-Corrected Thompson Sampling} (FC-TS), a Bayesian approach to non-stationary contextual bandits that transforms historical observations before incorporating them into the posterior. FC-TS estimates the drift online and applies observation-specific corrections that ``flow'' past rewards forward to the current time. Combined with adaptive weighting that reflects confidence in each correction, FC-TS achieves substantial improvements over existing methods when the environment exhibits structured non-stationarity.

\noindent \textbf{Contributions.} Our main contributions are as follows.
\begin{itemize}[nosep,leftmargin=1.5em]
    \item \textbf{Flow-corrected posterior updates for non-stationary bandits.} We propose a transport-based view of non-stationarity in linear contextual bandits, in which past rewards are corrected into present-time pseudo-observations and incorporated through confidence-weighted Bayesian updates rather than being uniformly discarded.

    \item \textbf{A modular algorithm with structured instantiations.} We develop \fcTS{}, a Thompson-sampling framework defined by a transport operator and a weighting rule, and instantiate it for (i) linear drift via online ridge regression and linear reward transport, (ii) periodic variation via phase-aware reuse across cycles, and (iii) recurring regimes via changepoint detection and regime-specific posterior memory, including an adaptive variant that combines trends and multiple periodicities.

    \item \textbf{Broad empirical evaluation and ablations.} We evaluate \fcTS{} across five controlled and semi-synthetic case studies, including linear drift, periodic variation, recurring regimes, compound non-stationarity, and portfolio selection, and use ablations to isolate the value of transport correction and structured memory.
\end{itemize}

\noindent \textbf{Paper Organization.} Section~\ref{sec:related} discusses related work. Section~\ref{sec:method} presents our methodology, including the problem formulation, algorithm, and drift-specific instantiations. Section~\ref{sec:experiments} provides experimental evaluation. Section~\ref{sec:conclusion} concludes with discussion and future directions.

\section{Related Work}
\label{sec:related}

\noindent \textbf{Non-stationary bandits and forgetting-based methods.} Sliding-window UCB \citep{garivier2011upper}, discounted UCB/TS \citep{kocsis2006discounted,raj2017taming}, and weighted linear bandits \citep{russac2019weighted} adapt to drift by down-weighting old observations uniformly with age. Restarting and variation-budget methods \citep{besbes2014stochastic,auer2019adaptively,cheung2019learning} achieve near-optimal rates under abrupt or budget-constrained change. \fcTS{} differs by applying \emph{observation-specific} corrections that preserve the value of old data when drift structure permits.

\noindent \textbf{Change-point detection and Bayesian bandits.} Detection-based approaches \citep{liu2018change,cao2019nearly,besson2022efficient} reset upon detecting a change; our regime instantiation extends this by \emph{retaining} regime-specific posteriors for reuse. Thompson sampling \citep{thompson1933likelihood,agrawal2013thompson,russo2014learning} provides our Bayesian backbone; \fcTS{} modifies how observations enter the posterior rather than how the posterior is maintained. 

%Recent bandit and safe-learning work further illustrates the breadth of sequential decision-making settings considered in this literature \citep{yifru2024concurrent,baheri2023llms,baheri2025multilevel}.

\noindent \textbf{Optimal transport.} The ``flow-corrected'' name draws inspiration from optimal transport \citep{villani2009optimal,peyre2019computational}; we transport historical \emph{reward observations} to align with the current reward distribution rather than solving a full OT problem.

% =====================================================================

\section{Method}
\label{sec:method}
% =====================================================================

We introduce \fcTS{} (Flow-Corrected Thompson Sampling), a Thompson sampling framework designed for contextual bandits in which the reward model changes over time but historical observations need not be discarded. The main difficulty is that old data are both valuable and stale: retaining them can reduce posterior variance, yet using them naively can bias the learner toward parameters that no longer describe the current environment. \fcTS{} addresses this tension by separating two roles that are often conflated in non-stationary bandits. First, a \emph{transport operator} rewrites past rewards as estimates of what they would have been under the current reward model. Second, a \emph{confidence weight} controls how much each transported observation should influence the posterior when the correction is uncertain.

This separation yields a common algorithmic template for several forms of non-stationarity. Linear trends are handled by estimating local drift and shifting old rewards forward in time; periodic structure is handled by reusing observations from nearby phases; recurring regimes are handled by storing and reloading regime-specific posteriors. In all cases, the action-selection rule remains Thompson sampling with a Gaussian working posterior, while the transport and weighting rules encode the assumed temporal structure.

% ---------------------------------------------------------------------
\subsection{Problem Setup and Design Principle}
\label{sec:setup}
% ---------------------------------------------------------------------

We consider a stochastic contextual bandit over $T$ rounds. At round $t$, the learner observes a context $x_t \in \reals^d$ with $\|x_t\|_2 \le 1$, selects an action $a_t \in [K] \coloneqq \{1,\dots,K\}$, and receives a reward.
\begin{equation}
\label{eq:reward}
    r_t = \langle w_t^{(a_t)}, x_t \rangle + \varepsilon_t,
    \qquad \varepsilon_t \sim \cN(0,\sigma^2),
\end{equation}
where $w_t^{(a)} \in \reals^d$ is the time-dependent parameter vector for action~$a$. We evaluate performance by cumulative regret
$\cR_T = \sum_{t=1}^{T}[\max_{a}\langle w_t^{(a)}, x_t\rangle - \langle w_t^{(a_t)}, x_t\rangle]$.

If the parameters were stationary, all past observations assigned to an action could be pooled in a standard Bayesian linear regression update. Non-stationarity breaks this pooling argument: for $s<t$, the reward $r_s$ is centered at $\langle w_s^{(a)},x_s\rangle$, whereas the current decision depends on $\langle w_t^{(a)},x_s\rangle$. The purpose of \fcTS{} is to recover as much of the statistical value of past data as possible while avoiding the systematic error induced by this mismatch.

At each round, \fcTS{} maintains a per-action Gaussian working model
\begin{equation}
\label{eq:posterior}
    w_t^{(a)} \mid \cH_t \;\sim\; \cN\!\bigl(\mu_t^{(a)},\; (\Lambda_t^{(a)})^{-1}\bigr),
\end{equation}
with prior $w_t^{(a)} \sim \cN(m_t^{(a)}, \lambda^{-1} I_d)$ and history $\cH_t = \{(x_s, a_s, r_s)\}_{s < t}$. In the practical algorithms below, we set $m_t^{(a)}=0$; the framework also permits time-dependent prior means when such side information is available.

% ---------------------------------------------------------------------
\subsection{Flow-Corrected Posterior Update}
\label{sec:transport}
% ---------------------------------------------------------------------

For each past observation $(x_s,a_s,r_s)$ and current time $t$, \fcTS{} constructs a transported reward
\begin{equation}
\label{eq:transported}
    \hat{r}_{s \to t} \;=\; r_s + \widehat{\Delta}_{s \to t}(x_s, a_s),
\end{equation}
where $\widehat{\Delta}_{s \to t}(x_s,a_s)$ estimates the shift in mean reward from time $s$ to time $t$. Ideally, this quantity approximates
$\langle w_t^{(a_s)} - w_s^{(a_s)}, x_s \rangle$, so that $\hat{r}_{s\to t}$ behaves like an observation aligned with the current parameter for action $a_s$.

Transport alone is not enough: an aggressive correction can introduce noise when the drift model is inaccurate. \fcTS{} therefore assigns each transported observation a confidence weight $\omega_{s,t}\in[0,1]$. A weight of one treats the transported sample as fully reliable, while smaller weights inflate its effective noise variance and reduce its influence. For each action $a$, the resulting precision matrix, information vector, and posterior mean are
\begin{align}
    \Lambda_t^{(a)} &= \lambda I_d + \sigma^{-2} \sum_{\substack{s < t \\ a_s = a}} \omega_{s,t}\, x_s x_s^\top,
    \label{eq:precision}\\[4pt]
    \eta_t^{(a)} &= \lambda\, m_t^{(a)} + \sigma^{-2} \sum_{\substack{s < t \\ a_s = a}} \omega_{s,t}\, \hat{r}_{s \to t}\, x_s,
    \label{eq:info_vector}\\[4pt]
    \mu_t^{(a)} &= \bigl(\Lambda_t^{(a)}\bigr)^{-1} \eta_t^{(a)}.
    \label{eq:mean}
\end{align}
This update makes explicit the bias--variance trade-off created by non-stationarity. The transported reward targets bias by aligning old observations with the present, whereas the confidence weight targets variance and model misspecification by controlling how strongly each corrected sample enters the regression problem.

Action selection is then standard Thompson sampling under the working model. At round $t$, the learner samples
$\tilde{w}_t^{(a)} \sim \cN\!\bigl(\mu_t^{(a)}, (\Lambda_t^{(a)})^{-1}\bigr)$
for each action and chooses $a_t = \argmax_{a \in [K]} \langle \tilde{w}_t^{(a)}, x_t \rangle$. \Cref{alg:fcts} summarizes the generic template. The only components that change across environments are the transport operator $\widehat{\Delta}$ and the weighting rule $\omega$.

\begin{algorithm}[t]
\caption{\fcTS{} (generic template)}
\label{alg:fcts}
\begin{algorithmic}[1]
\REQUIRE Prior precision $\lambda$, noise variance $\sigma^2$, prior mean rule $m$, transport operator $\widehat{\Delta}$, weight rule $\omega$
\STATE Initialize per-action working models $\{(\mu^{(a)}, \Lambda^{(a)})\}_{a=1}^{K}$ from the prior
\STATE Initialize the state required by the chosen transport model
\FOR{$t = 1, 2, \dots, T$}
    \STATE Observe context $x_t$
    \FOR{$a = 1, \dots, K$}
        \STATE Build $\Lambda_t^{(a)}$, $\eta_t^{(a)}$, and $\mu_t^{(a)}$ using \eqref{eq:precision}--\eqref{eq:mean}
        \STATE Sample $\tilde{w}_t^{(a)} \sim \cN\!\bigl(\mu_t^{(a)}, (\Lambda_t^{(a)})^{-1}\bigr)$
    \ENDFOR
    \STATE Select $a_t = \argmax_a \langle \tilde{w}_t^{(a)}, x_t \rangle$ and observe reward $r_t$
    \STATE Update the transport model state from the new observation
    \STATE Store $(x_t, a_t, r_t)$ in memory, subject to the chosen truncation rule
\ENDFOR
\end{algorithmic}
\end{algorithm}

% =====================================================================
\subsection{Practical Instantiations}
\label{sec:instantiations}
% =====================================================================

The template above is useful only if the transport and weighting rules can be specified in a way that matches the structure of the environment. We study three cases that cover common non-stationary patterns: smooth linear drift, periodic variation, and recurring regimes. These cases share the same posterior update but differ in how they decide which historical observations remain comparable to the present.

\subsubsection{Linear drift.}
\label{sec:linear_inst}
In the linear-drift setting, each action parameter is assumed to evolve approximately as $w_t^{(a)} \approx \theta^{(a)} + t\,\delta^{(a)}$, where $\delta^{(a)}\in\reals^d$ is an action-specific drift rate. The goal is not to estimate a global model once, but to maintain a local estimate of the current trend and use it to shift earlier rewards forward.

Given an online estimate $\hat{\delta}_t^{(a)}$, \fcTS{} transports an observation from time $s$ to time $t$ by
\begin{equation}
\label{eq:linear_transport}
    \hat{r}_{s \to t} = r_s + (t - s)\langle \hat{\delta}_t^{(a)}, x_s \rangle.
\end{equation}
The drift estimate is obtained by ridge regression on a rolling window of size $H_\delta$. Let $\cI_t^{(a)} = \{s \in [t - H_\delta, t-1] : a_s = a\}$, let $\bar{s}$ be the mean time index in the window, and define the augmented feature $\phi_s = [x_s;\; (s - \bar{s})x_s] \in \reals^{2d}$. We estimate the local intercept and slope by
\begin{equation}
\label{eq:drift_ridge}
    \begin{bmatrix} \hat{\theta}_t^{(a)} \\ \hat{\delta}_t^{(a)} \end{bmatrix}
    = \argmin_{\theta,\delta}
    \sum_{s \in \cI_t^{(a)}}
    \bigl(r_s - \langle \theta, x_s \rangle - (s - \bar{s})\langle \delta, x_s \rangle\bigr)^2
    + \lambda_\delta(\|\theta\|_2^2 + \|\delta\|_2^2).
\end{equation}
Because extrapolation becomes less reliable over longer horizons, the transported samples are weighted by exponential decay,
\begin{equation}
\label{eq:linear_weight}
    \omega_{s,t} = \gamma^{t-s}, \qquad \gamma\in(0,1).
\end{equation}
This design lets nearby observations enter with high confidence while still allowing older observations to contribute when the drift correction makes them useful.

\subsubsection{Periodic variation.}
\label{sec:periodic_inst}
Some environments are not monotonic in time but are recurring. When rewards follow a known period $P$, observations from the same phase can be more relevant than more recent observations from a different phase. We model this case by writing $w_t^{(a)} \approx w_{\mathrm{ph}(t)}^{(a)}$ with $\mathrm{ph}(t)=t \bmod P$. Since samples from the same phase are already aligned, the transported reward is simply $\hat{r}_{s\to t}=r_s$.

The confidence weight is determined by phase proximity. Let
$d_P(s,t) = \min(|\mathrm{ph}(s)-\mathrm{ph}(t)|, P-|\mathrm{ph}(s)-\mathrm{ph}(t)|)$
be the circular distance between phases. We use
\begin{equation}
\label{eq:periodic_weight}
    \omega_{s,t}
    = \exp\!\Bigl(-\frac{d_P(s,t)^2}{2\tau^2}\Bigr)
    \cdot \gamma_{\mathrm{cyc}}^{\lfloor |t - s| / P \rfloor},
\end{equation}
where $\tau$ controls the phase bandwidth and $\gamma_{\mathrm{cyc}}\in(0,1]$ allows mild decay across cycles. This creates a phase memory: old observations can remain influential when they occur at the right part of the cycle. For environments with multiple periods $\{P_j\}$, we combine the corresponding kernels multiplicatively,
\begin{equation}
\label{eq:multi_periodic_weight}
    \omega_{s,t}^{(\mathrm{periodic})}
    = \prod_j
    \exp\!\left(-\frac{d_{P_j}(s,t)^2}{2\tau_j^2}\right)
    \gamma_{\mathrm{cyc},j}^{\lfloor |t-s|/P_j \rfloor}.
\end{equation}

\subsubsection{Recurring regimes.}
\label{sec:regime_inst}
A third pattern arises when the environment switches among a finite set of regimes, with the possibility that a regime observed earlier may return later. In this case, the main risk is not gradual drift but forgetting useful evidence from a regime that temporarily disappears. \fcTS{} handles this setting with a bank of regime-specific working posteriors $\{(\mu^{(a,k)},\Lambda^{(a,k)})\}_{a,k}$, one for each action $a$ and regime $k$.

When regime $k_t$ is active, only observations assigned to that regime are used in the current update:
\begin{equation}
\label{eq:regime_weight}
    \omega_{s,t}=\mathbb{I}\{k_s=k_t\},
    \qquad
    \hat{r}_{s\to t}=r_s.
\end{equation}
Thus, a recurring regime can be re-entered with its previously accumulated posterior rather than starting from scratch. To identify switches, we monitor the prediction residual $e_t = r_t - \langle \mu_{t-1}^{(a_t)}, x_t \rangle$ and apply a CUSUM test to $e_t^2$. When the statistic exceeds a threshold $h$, recent observations are scored under the stored regimes using marginal likelihood. The best-scoring regime is reloaded if its score is sufficiently high; otherwise, a new regime is created.

\subsubsection{Composition and computational cost.}
\label{sec:adaptive}
The same interface also supports a composed structure. For example, when linear trend and seasonality coexist, we use linear transport together with periodic weighting:
\begin{equation}
\label{eq:composed_transport}
    \hat{r}_{s \to t} = r_s + (t-s)\langle \hat{\delta}_t^{(a)}, x_s \rangle,
    \qquad
    \omega_{s,t} = \gamma^{t-s}\,\omega_{s,t}^{(\mathrm{periodic})}.
\end{equation}
This composition is possible because the posterior update only requires a transported response and a confidence weight; it does not require a separate algorithm for each drift pattern.

%With exponential weighting, memory can be truncated once $\omega_{s,t}<\omega_{\min}$. The resulting active history size is $O(\log(\omega_{\min}^{-1})/\log(\gamma^{-1}))$, independent of $T$. Maintaining the sufficient statistics $(\Lambda^{(a)},\eta^{(a)})$ with rank-one Cholesky updates gives $O(d^2)$ cost per action per round.

% =====================================================================
% APPENDIX
% =============================================================

\section{Experiments}
\label{sec:experiments}

We evaluate \fcTS{} on five case studies of increasing complexity: linear parameter drift, periodic variation, recurring regime switches, compound (drift + changepoints), and a semi-synthetic portfolio-selection task with multiple overlapping non-stationarities. All experiments report cumulative regret $\mathcal{R}_T = \sum_{t=1}^{T}\bigl[\max_a \mu_t(a, x_t) - \mu_t(a_t, x_t)\bigr]$ averaged over multiple seeds with $\pm 1$ standard deviation. Full per-case tables, learning curves, and additional diagnostics are in \cref{app:experiments_details}.

\paragraph{Baselines.}
We compare four representative non-stationary Thompson-sampling variants: \stdTS{} (standard TS), \swTS{} (sliding-window TS, $W=200$), \dTS{} (discounted TS, $\gamma=0.995$), and \rTS{} (restarting TS, $\tau=400$). All share the same Bayesian linear-regression backbone with Gaussian likelihood ($\sigma = 0.1$ synthetic, $\sigma = 0.15$ portfolio). Baseline hyperparameters were tuned via grid search to favor the strongest competitor in each setting. \fcTS{} uses online ridge regression for drift estimation ($\lambda_\delta = 1.0$, window 200, warm-up 50), with $\gamma = 0.999$ for synthetic experiments and $\gamma = 0.9995$ for the portfolio task.

\paragraph{Headline results.}
Table~\ref{tab:summary} consolidates the results: \fcTS{} achieves the lowest cumulative regret in every setting, with improvements of \textbf{14.1\%--59.0\%} over the strongest baseline (which itself varies by setting). Notably, no single forgetting-based baseline dominates across environments: \swTS{} wins under linear drift and compound non-stationarity, \stdTS{} under pure periodicity (since the time-averaged parameter equals $w_0$), \rTS{} under regimes, and \dTS{} on the portfolio task. \fcTS{} is consistently best because it adapts its data-reuse strategy to the structure of the non-stationarity rather than committing to one forgetting rule.

\begin{table}[h]
\centering
\caption{Summary across all five case studies: \fcTS{} vs.\ the best baseline (varying by setting). Per-case full tables in \cref{app:experiments_details}.}
\label{tab:summary}
\vspace{0.3em}
\small
\begin{tabular}{lccccc}
\toprule
& \textbf{Linear} & \textbf{Periodic} & \textbf{Regime} & \textbf{Compound} & \textbf{Portfolio} \\
& (Case 1) & (Case 2) & (Case 3) & (Case 4) & (Case 5) \\
\midrule
$(K,d,T)$            & $(5,10,2k)$ & $(5,10,2k)$ & $(5,10,2k)$ & $(5,10,3k)$ & $(6,12,3k)$ \\
Seeds                & 10 & 10 & 10 & 10 & 5 \\
\midrule
Best baseline        & \swTS & \stdTS & \rTS & \swTS & \dTS \\
Baseline regret      & 51.0 & 344.1 & 140.6 & 385.7 & 115.2 \\
\fcTS{} regret       & \best{43.8} & \best{172.2} & \best{57.6} & \best{272.0} & \best{82.8} \\
\midrule
\textbf{Improvement} & \textbf{14.1\%} & \textbf{50.0\%} & \textbf{59.0\%} & \textbf{29.5\%} & \textbf{28.1\%} \\
\bottomrule
\end{tabular}
\end{table}

\paragraph{Per-case observations.}
The pattern across cases reveals when transport-based reuse pays off most.
\emph{(Case 1, Linear drift, $w_t^{(a)} = w_0^{(a)} + t\delta^{(a)}$ with $\delta^{(a)} \sim \mathcal{N}(0, 0.003^2 I)$):} \fcTS{} matches \swTS{} during the first $\sim$500 rounds of drift warm-up, then progressively separates, winning 8 of 10 seeds.
\emph{(Case 2, Periodic, sinusoidal with $P=200$):} the largest synthetic gain (50.0\%) comes from the \emph{phase-memory} effect: by the fifth cycle, \fcTS{} has accumulated four previous cycles of same-phase data per bin, yielding per-cycle regret that drops from $\sim$70 to $\sim$4 over 10 cycles, while baselines plateau at $\sim$30–35.
\emph{(Case 3, Three recurring regimes):} regime-specific posterior reloading on changepoint detection (median delay $<$ 15 rounds) gives the largest improvement (59.0\%); \rTS{}'s periodic restarts only sometimes align with regime boundaries.
\emph{(Case 4, Compound, linear drift between true changepoints at $t \in \{700, 1500\}$):} the detector fires with a 1–3 round delay; drift correction handles smooth evolution between changes while reset handles abrupt ones, yielding a 29.5\% improvement.
\emph{(Case 5, Semi-synthetic portfolio, $K=6$ strategies over $T=3000$ trading days, monthly + quarterly cycles and growth-to-value drift):} \fcTS{}-adaptive (linear + periodic) achieves a 28.1\% improvement; an ablation (\cref{app:experiments_details}, Fig.~\ref{fig:portfolio}) shows drift correction is the dominant component (27.3\% alone) while periodic awareness provides a smaller additive benefit, reflecting that the long-horizon rotation dominates the short monthly cycle.

\begin{figure}[t]
\centering
\begin{subfigure}[t]{0.32\textwidth}
    \includegraphics[width=\textwidth]{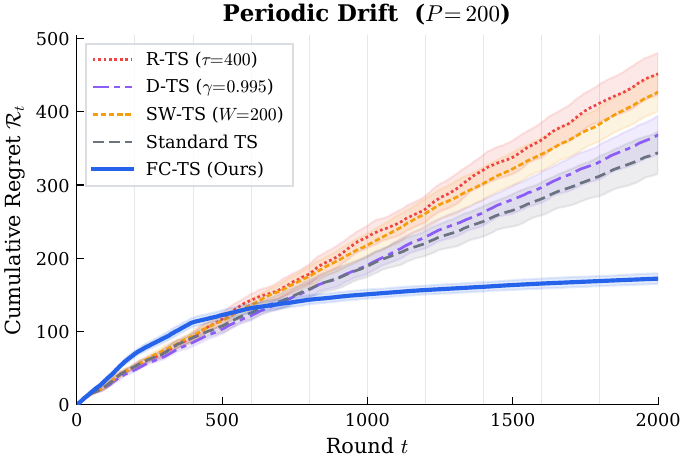}
    \caption{Case 2: Periodic.}
    \label{fig:main_periodic}
\end{subfigure}
\hfill
\begin{subfigure}[t]{0.32\textwidth}
    \includegraphics[width=\textwidth]{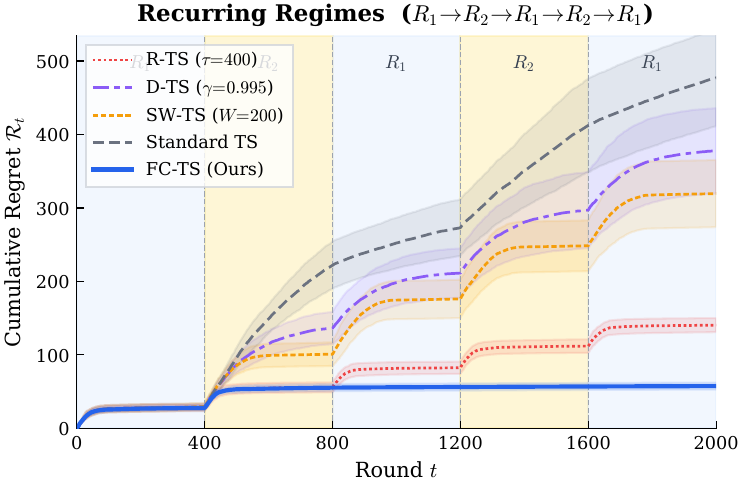}
    \caption{Case 3: Recurring regimes.}
    \label{fig:main_regime}
\end{subfigure}
\hfill
\begin{subfigure}[t]{0.32\textwidth}
    \includegraphics[width=\textwidth]{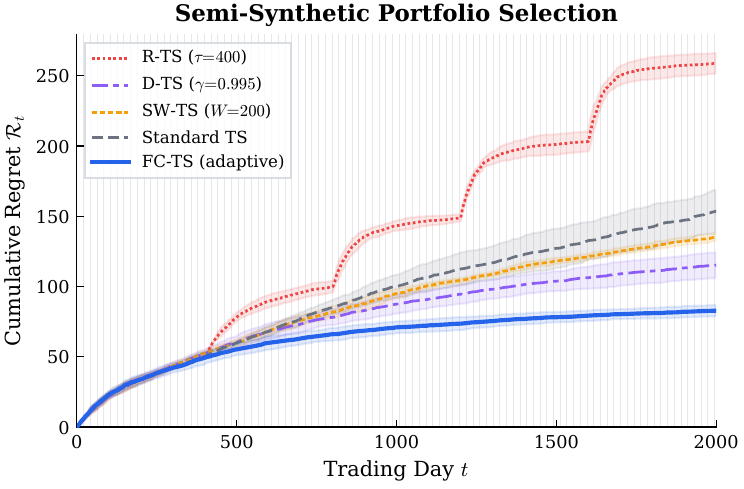}
    \caption{Case 5: Portfolio.}
    \label{fig:main_portfolio}
\end{subfigure}
\caption{Cumulative regret curves for three representative case studies. \fcTS{} (blue, solid) is consistently below all baselines. Additional cases and instantaneous-regret/per-cycle/per-segment diagnostics are in \cref{app:experiments_details}.}
\label{fig:main_results}
\end{figure}

\paragraph{Ablations.}
Two ablation studies in \cref{app:ablation} isolate the contribution of the transport correction itself. (i) Holding effective sample size fixed via $\gamma$, the no-transport variant has the \emph{highest} regret while \fcTS{} has the lowest: more data hurts without transport correction but helps with it (\cref{app:neff}). (ii) Sweeping the drift scale $\delta \in [0.0005, 0.012]$, the transport correction provides a 47--57\% regret reduction over the no-transport ablation at every drift level (\cref{app:transport}). Together these confirm that the transport operator, not the choice of $\gamma$ or posterior-rebuild mechanism, drives \fcTS{}'s gains.

\section{Conclusion}
\label{sec:conclusion}

We introduced \emph{Flow-Corrected Thompson Sampling} (\fcTS{}), a Bayesian framework for non-stationary linear contextual bandits that treats historical data as reusable evidence rather than obsolete information. Its core idea is to convert past interactions into present-time pseudo-observations by transporting rewards under a drift model and weighting them by transport confidence. This provides a unified template for linear drift, periodic variation, and recurring regimes through online correction, phase-aware reuse, and regime-specific posterior memory. Empirically, \fcTS{} consistently improves over forgetting-based baselines across synthetic and semi-synthetic benchmarks, especially when temporal structure recurs. These results suggest that structured non-stationarity is better addressed by correcting and reusing experience than by uniformly discarding it. Future work includes learning drift structure online with calibrated uncertainty, extending \fcTS{} beyond linear rewards to generalized linear or neural bandits, testing robustness to transport-estimation error, and developing safeguards for high-stakes settings where unmodeled distribution shifts and heavy-tailed noise may arise.

%%%%%%%%%%%%%%%%%%%%%%%%%%%%%%%%%%%%%%%%%%%%%%%%%%%%%%%%%%%%%%%%
%% Appendices
%%%%%%%%%%%%%%%%%%%%%%%%%%%%%%%%%%%%%%%%%%%%%%%%%%%%%%%%%%%%%%%%
%\appendix

%%%%%%%%%%%%%%%%%%%%%%%%%%%%%%%%%%%%%%%%%%%%%%%%%%%%%%%%%%%%%%%%
%% NOTE: THIS MARKS THE END OF THE "MAIN TEXT"
%%%%%%%%%%%%%%%%%%%%%%%%%%%%%%%%%%%%%%%%%%%%%%%%%%%%%%%%%%%%%%%%

%%%%%%%%%%%%%%%%%%%%%%%%%%%%%%%%%%%%%%%%%%%%%%%%%%%%%%%%%%%%%%%%
%% Bibliography
%%%%%%%%%%%%%%%%%%%%%%%%%%%%%%%%%%%%%%%%%%%%%%%%%%%%%%%%%%%%%%%%
\newpage
\bibliography{main}

@inproceedings{bouneffouf2020survey,
  title={Survey on Applications of Multi-Armed and Contextual Bandits},
  author={Bouneffouf, Djallel and Rish, Irina and Aggarwal, Charu},
  booktitle={2020 IEEE Congress on Evolutionary Computation (CEC)},
  pages={1--8},
  year={2020},
  organization={IEEE},
  doi={10.1109/CEC48606.2020.9185782}
}

@article{afsar2022reinforcement,
  title={Reinforcement Learning Based Recommender Systems: A Survey},
  author={Afsar, M. Mehdi and Crump, Trafford and Far, Behrouz},
  journal={ACM Computing Surveys},
  volume={55},
  number={7},
  pages={1--38},
  year={2022},
  publisher={ACM},
  doi={10.1145/3543846}
}

@article{razzaghi2024survey,
  title={A Survey on Reinforcement Learning in Aviation Applications},
  author={Razzaghi, Pouria and Tabrizian, Amin and Guo, Wei and Chen, Shulu and Taye, Abenezer and Thompson, Ellis and Bregeon, Alexis and Baheri, Ali and Wei, Peng},
  journal={Engineering Applications of Artificial Intelligence},
  volume={136},
  pages={108911},
  year={2024},
  publisher={Elsevier},
  doi={10.1016/j.engappai.2024.108911}
}

@article{kober2013reinforcement,
  title={Reinforcement Learning in Robotics: A Survey},
  author={Kober, Jens and Bagnell, J. Andrew and Peters, Jan},
  journal={The International Journal of Robotics Research},
  volume={32},
  number={11},
  pages={1238--1274},
  year={2013},
  publisher={SAGE Publications},
  doi={10.1177/0278364913495721}
}

@inproceedings{garivier2011upper,
  title={On Upper-Confidence Bound Policies for Switching Bandit Problems},
  author={Garivier, Aur{\'e}lien and Moulines, Eric},
  booktitle={Algorithmic Learning Theory},
  pages={174--188},
  year={2011},
  publisher={Springer},
  doi={10.1007/978-3-642-24412-4_16}
}

@inproceedings{besbes2014stochastic,
  title={Stochastic Multi-Armed-Bandit Problem with Non-Stationary Rewards},
  author={Besbes, Omar and Gur, Yonatan and Zeevi, Assaf},
  booktitle={Advances in Neural Information Processing Systems},
  volume={27},
  pages={199--207},
  year={2014},
  url={https://proceedings.neurips.cc/paper/2014/hash/91ba7292e5388b90b58d0b839a7f19ec-Abstract.html}
}

@inproceedings{cheung2019learning,
  title={Learning to Optimize under Non-Stationarity},
  author={Cheung, Wang Chi and Simchi-Levi, David and Zhu, Ruihao},
  booktitle={Proceedings of the Twenty-Second International Conference on Artificial Intelligence and Statistics},
  pages={1079--1087},
  volume={89},
  year={2019},
  organization={PMLR},
  url={https://proceedings.mlr.press/v89/cheung19b.html}
}

@inproceedings{auer2019adaptively,
  title={Adaptively Tracking the Best Bandit Arm with an Unknown Number of Distribution Changes},
  author={Auer, Peter and Gajane, Pratik and Ortner, Ronald},
  booktitle={Proceedings of the Thirty-Second Conference on Learning Theory},
  pages={138--158},
  volume={99},
  year={2019},
  organization={PMLR},
  url={https://proceedings.mlr.press/v99/auer19a.html}
}

@inproceedings{kocsis2006discounted,
  title={Discounted {UCB}},
  author={Kocsis, Levente and Szepesv{\'a}ri, Csaba},
  booktitle={2nd PASCAL Challenges Workshop},
  volume={2},
  year={2006}
}

@article{raj2017taming,
  title={Taming Non-Stationary Bandits: A Bayesian Approach},
  author={Raj, Vishnu and Kalyani, Sheetal},
  journal={arXiv preprint arXiv:1707.09727},
  year={2017},
  eprint={1707.09727},
  archivePrefix={arXiv},
  primaryClass={cs.LG},
  doi={10.48550/arXiv.1707.09727},
  url={https://arxiv.org/abs/1707.09727}
}

@inproceedings{russac2019weighted,
  title={Weighted linear bandits for non-stationary environments},
  author={Russac, Yoan and Vernade, Claire and Capp{\'e}, Olivier},
  booktitle={Advances in Neural Information Processing Systems},
  volume={32},
  year={2019}
}

@inproceedings{liu2018change,
  title={A Change-Detection Based Framework for Piecewise-Stationary Multi-Armed Bandit Problem},
  author={Liu, Fang and Lee, Joohyun and Shroff, Ness B.},
  booktitle={Proceedings of the Thirty-Second AAAI Conference on Artificial Intelligence},
  pages={3651--3658},
  year={2018},
  publisher={AAAI Press},
  doi={10.1609/aaai.v32i1.11746}
}

@inproceedings{cao2019nearly,
  title={Nearly Optimal Adaptive Procedure with Change Detection for Piecewise-Stationary Bandit},
  author={Cao, Yang and Wen, Zheng and Kveton, Branislav and Xie, Yao},
  booktitle={Proceedings of the Twenty-Second International Conference on Artificial Intelligence and Statistics},
  pages={418--427},
  volume={89},
  year={2019},
  organization={PMLR},
  url={https://proceedings.mlr.press/v89/cao19a.html}
}

@article{besson2022efficient,
  title={Efficient change-point detection for tackling piecewise-stationary bandits},
  author={Besson, Lilian and Kaufmann, Emilie and Maillard, Odalric-Ambrym and Seznec, Julien},
  journal={Journal of Machine Learning Research},
  volume={23},
  number={77},
  pages={1--40},
  year={2022}
}

@article{thompson1933likelihood,
  title={On the Likelihood that One Unknown Probability Exceeds Another in View of the Evidence of Two Samples},
  author={Thompson, William R.},
  journal={Biometrika},
  volume={25},
  number={3/4},
  pages={285--294},
  year={1933},
  publisher={Oxford University Press},
  doi={10.1093/biomet/25.3-4.285}
}

@inproceedings{agrawal2013thompson,
  title={Thompson Sampling for Contextual Bandits with Linear Payoffs},
  author={Agrawal, Shipra and Goyal, Navin},
  booktitle={Proceedings of the 30th International Conference on Machine Learning},
  pages={127--135},
  volume={28},
  year={2013},
  organization={PMLR},
  url={https://proceedings.mlr.press/v28/agrawal13.html}
}

@inproceedings{russo2014learning,
  title={Learning to optimize via information-directed sampling},
  author={Russo, Daniel and Van Roy, Benjamin},
  booktitle={Advances in Neural Information Processing Systems},
  volume={27},
  year={2014}
}

@book{villani2009optimal,
  title={Optimal Transport: Old and New},
  author={Villani, C{\'e}dric},
  series={Grundlehren der Mathematischen Wissenschaften},
  volume={338},
  year={2009},
  publisher={Springer},
  doi={10.1007/978-3-540-71050-9}
}

@article{peyre2019computational,
  title={Computational Optimal Transport},
  author={Peyr{\'e}, Gabriel and Cuturi, Marco},
  journal={Foundations and Trends in Machine Learning},
  volume={11},
  number={5--6},
  pages={355--607},
  year={2019},
  doi={10.1561/2200000073}
}

@article{baheri2023llms,
  title={{LLM}s-augmented Contextual Bandit},
  author={Baheri, Ali and Alm, Cecilia O.},
  journal={arXiv preprint arXiv:2311.02268},
  note={Foundation Models for Decision Making workshop, NeurIPS 2023},
  year={2023},
  eprint={2311.02268},
  archivePrefix={arXiv},
  primaryClass={cs.LG},
  doi={10.48550/arXiv.2311.02268},
  url={https://arxiv.org/abs/2311.02268}
}
\bibliographystyle{rlj}

%%%%%%%%%%%%%%%%%%%%%%%%%%%%%%%%%%%%%%%%%%%%%%%%%%%%%%%%%%%%%%%%
% AUTHOR: If your paper has no supplementary materials, you may 
%         comment out the line below, which creates the title for
%         the supplementary materials.
%%%%%%%%%%%%%%%%%%%%%%%%%%%%%%%%%%%%%%%%%%%%%%%%%%%%%%%%%%%%%%%%
\beginSupplementaryMaterials
% 

% =====================================================================
% APPENDIX
% =============================================================
\appendix

\section{Detailed Experimental Results}
\label{app:experiments_details}

%This appendix presents the full per-case-study tables, learning curves, and additional diagnostics that were summarized in \cref{sec:experiments} of the main paper. 
\subsection{Setup (full)}
\label{app:setup_full}

We evaluate \fcTS{} across five case studies of increasing complexity, followed by a systematic robustness analysis.
The first three experiments use synthetic environments where the drift structure is fully specified, enabling precise regret computation.
The fourth introduces compound non-stationarity (simultaneous drift and changepoints), and the fifth is a semi-synthetic financial portfolio selection task with realistic market dynamics.
All experiments report cumulative regret $\mathcal{R}_T = \sum_{t=1}^{T} \bigl[\max_a \mu_t(a, x_t) - \mu_t(a_t, x_t)\bigr]$ averaged over independent seeds, with $\pm 1$ standard deviation bands.

\paragraph{Baselines.}
We compare four representative non-stationary Thompson Sampling variants:
\begin{itemize}[nosep,leftmargin=1.5em]
    \item \textbf{\stdTS}: Standard Thompson Sampling with no non-stationarity handling.
    \item \textbf{\swTS} ($W\!=\!200$): Sliding-window TS that maintains a posterior using only the most recent $W$ observations.
    \item \textbf{\dTS} ($\gamma\!=\!0.995$): Discounted TS that exponentially down-weights older observations.
    \item \textbf{\rTS} ($\tau\!=\!400$): Restarting TS that periodically resets the posterior every $\tau$ rounds.
\end{itemize}

\noindent
All methods share the same Bayesian linear regression backbone with Gaussian likelihood ($\sigma = 0.1$ for synthetic tasks, $\sigma = 0.15$ for the portfolio task).
Baseline hyperparameters were tuned via grid search to favor the strongest competitor in each setting. \fcTS{} uses online ridge regression for drift estimation (regularization $\lambda_\delta = 1.0$) with a sliding estimation window of 200 observations and a warm-up period of 50 rounds.
The discount factor is $\gamma = 0.999$ for all synthetic experiments and $\gamma = 0.9995$ for the longer-horizon portfolio task. 

%All experiments are reproducible via fixed random seeds, and source code accompanies this paper.

% ══════════════════════════════════════════════════════════════════
\noindent \textbf{Case Study 1: Linear Parameter Drift}

\noindent \textbf{Setup.}
We consider $K\!=\!5$ arms with $d\!=\!10$-dimensional contexts over $T\!=\!2{,}000$ rounds.
The reward parameter for each arm drifts linearly: $w_t^{(a)} = w_0^{(a)} + t \cdot \delta^{(a)}$, where $\delta^{(a)} \sim \mathcal{N}(0, 0.003^2 I)$.
This is the setting where \fcTS{}'s linear transport operator is exactly specified, serving as a best-case validation.

\noindent \textbf{Results.}
Table~\ref{tab:linear} reports the final cumulative regret over 10 seeds.
\fcTS{} achieves $\mathcal{R}_T = 43.8 \pm 12.9$, a \textbf{14.1\%} improvement over the best baseline (\swTS{} at $51.0 \pm 4.9$).
The advantage is consistent: \fcTS{} wins in 8 of 10 seeds, with the two losses occurring when stochastic drift directions happen to align favorably with \swTS{}'s fixed window.

\begin{table}[h]
\centering
\caption{Final cumulative regret $\mathcal{R}_T$ on the linear drift environment ($K\!=\!5$, $d\!=\!10$, $T\!=\!2{,}000$, 10 seeds). Best result in \best{blue bold}.}
\label{tab:linear}
\vspace{0.5em}
\begin{tabular}{lcc}
\toprule
\textbf{Algorithm} & \textbf{Regret} (mean $\pm$ std) & \textbf{vs.\ best baseline} \\
\midrule
\stdTS                         & $166.7 \pm 37.2$ & --- \\
\swTS{} ($W\!=\!200$)         & $51.0 \pm 4.9$   & (best baseline) \\
\dTS{} ($\gamma\!=\!0.995$)   & $67.2 \pm 6.1$   & --- \\
\rTS{} ($\tau\!=\!400$)       & $393.6 \pm 38.0$ & --- \\
\midrule
\fcTS{} (Ours)                & \best{$43.8 \pm 12.9$} & \imp{$-$14.1} \\
\bottomrule
\end{tabular}
\end{table}

\begin{figure}[t]
\centering
\begin{subfigure}[t]{0.48\textwidth}
    \includegraphics[width=\textwidth]{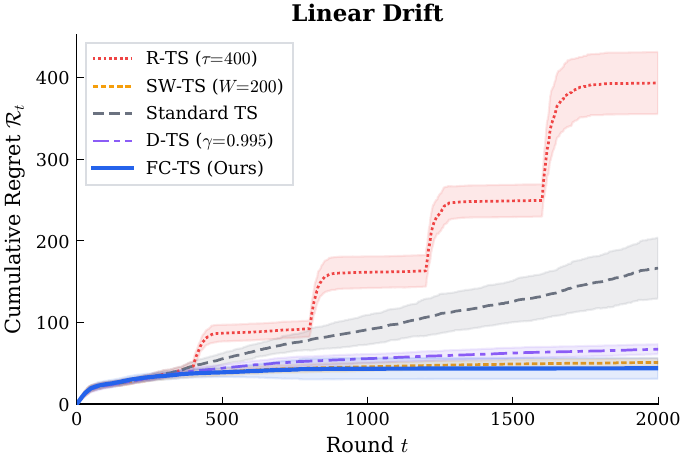}
    \caption{Cumulative regret $\mathcal{R}_t$ vs.\ round $t$.}
    \label{fig:linear_cumulative}
\end{subfigure}
\hfill
\begin{subfigure}[t]{0.48\textwidth}
    \includegraphics[width=\textwidth]{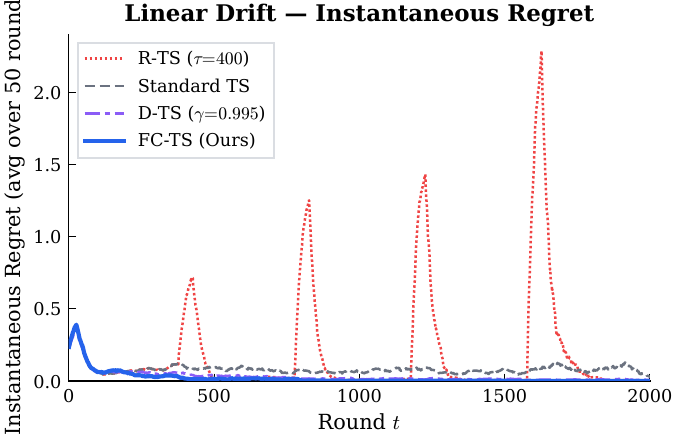}
    \caption{Instantaneous regret (50-round moving average).}
    \label{fig:linear_instant}
\end{subfigure}
\caption{\textbf{Case Study 1: Linear Drift.} \fcTS{} (blue, solid) achieves the lowest cumulative regret by transporting past observations forward along the estimated drift direction. \rTS{} shows periodic spikes at restart boundaries ($\tau = 400$). Shaded bands show $\pm 1$ std over 10 seeds.}
\label{fig:linear}
\end{figure}

The cumulative regret curves (Figure~\ref{fig:linear_cumulative}) reveal that \fcTS{} and \swTS{} track closely for the first $\sim$500 rounds while drift estimates warm up, after which \fcTS{} progressively separates.
\rTS{} performs the worst due to periodic information destruction at restart boundaries, visible as regret spikes every 400 rounds in the instantaneous regret plot (Figure~\ref{fig:linear_instant}).
Notably, \dTS{} underperforms \swTS{} here because exponential discounting cannot fully compensate for the accumulating bias of linear drift; it down-weights old data but does not correct it.

% ══════════════════════════════════════════════════════════════════
\noindent \textbf{Case Study 2: Periodic Parameter Variation}
\label{sec:periodic}
% ══════════════════════════════════════════════════════════════════

\noindent \textbf{Setup.}
Reward parameters oscillate sinusoidally with period $P\!=\!200$: $w_t^{(a)} = w_0^{(a)} + A^{(a)} \sin(2\pi t / P + \phi^{(a)})$, where amplitudes $A^{(a)}$ and phases $\phi^{(a)}$ are drawn randomly.
This is the canonical use case for \fcTS{}'s phase-binned posterior mechanism: observations from the same phase in previous cycles are directly reusable because the reward parameters are identical (up to noise).
We use $K\!=\!5$, $d\!=\!10$, $T\!=\!2{,}000$, and 10 seeds.

\noindent \textbf{Results.}
\fcTS{} achieves a dramatic \textbf{50.0\%} reduction in regret relative to the best baseline (Table~\ref{tab:periodic}).

\begin{table}[h]
\centering
\caption{Final cumulative regret on the periodic drift environment ($P\!=\!200$, 10 seeds).}
\label{tab:periodic}
\vspace{0.5em}
\begin{tabular}{lcc}
\toprule
\textbf{Algorithm} & \textbf{Regret} (mean $\pm$ std) & \textbf{vs.\ best baseline} \\
\midrule
\stdTS                         & $344.1 \pm 28.8$ & (best baseline) \\
\dTS{} ($\gamma\!=\!0.995$)   & $368.4 \pm 26.2$ & --- \\
\swTS{} ($W\!=\!200$)         & $427.1 \pm 26.1$ & --- \\
\rTS{} ($\tau\!=\!400$)       & $451.8 \pm 28.8$ & --- \\
\midrule
\fcTS{} (Ours)                & \best{$172.2 \pm 7.2$} & \imp{$-$50.0} \\
\bottomrule
\end{tabular}
\end{table}

\begin{figure}[t]
\centering
\begin{subfigure}[t]{0.48\textwidth}
    \includegraphics[width=\textwidth]{periodic_fig1_cumulative_regret.pdf}
    \caption{Cumulative regret. Vertical lines mark period boundaries.}
    \label{fig:periodic_cumulative}
\end{subfigure}
\hfill
\begin{subfigure}[t]{0.48\textwidth}
    \includegraphics[width=\textwidth]{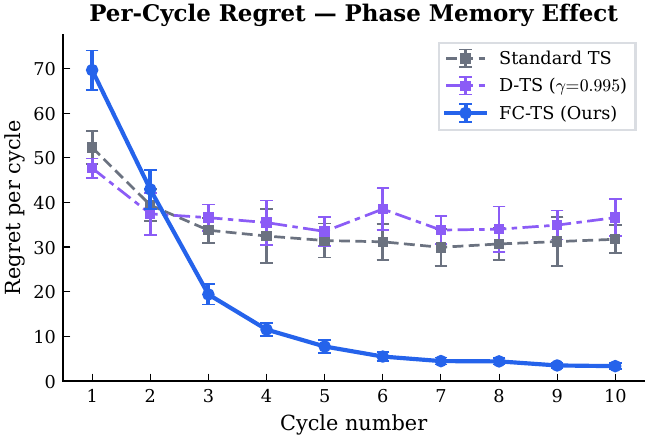}
    \caption{Per-cycle regret showing the phase memory effect.}
    \label{fig:periodic_phase}
\end{subfigure}
\caption{\textbf{Case Study 2: Periodic Drift.} \fcTS{} exploits phase recurrence by reusing same-phase observations across cycles. Per-cycle regret (right) drops to near zero after 5 cycles while baselines plateau.}
\label{fig:periodic}
\end{figure}

The performance gap is qualitatively explained by the \emph{phase memory} effect (Figure~\ref{fig:periodic}).
By the fifth cycle, \fcTS{} has accumulated 4 previous cycles of same-phase data per bin, producing posteriors that are approximately $5\times$ tighter than those of any baseline operating from scratch. The per cycle regret analysis (Figure~\ref{fig:periodic_phase}) confirms this: \fcTS{}'s per cycle regret drops from $\sim$70 (cycle 1) to $\sim$4 (cycle 10), while all baselines plateau around 30 to 35 per cycle. Interestingly, \stdTS{} is the strongest baseline here, not \swTS{} or \dTS{}.
This occurs because, under pure periodicity with no net drift, the time-averaged mean of $w_t^{(a)}$ equals $w_0^{(a)}$, so \stdTS{}'s posterior converges (slowly) to the correct time-averaged model.
In contrast, \swTS{} and \dTS{} actively discard useful older data.

% ══════════════════════════════════════════════════════════════════
\noindent \textbf{Case Study 3: Recurring Regime Switches}
\label{sec:regime}
% ══════════════════════════════════════════════════════════════════

\noindent \textbf{Setup.}
The environment alternates between 3 regimes, each with distinct reward parameters, in a pattern that recurs over $T\!=\!2{,}000$ rounds.
When a regime is revisited, its parameters are identical to the previous visit.
This tests \fcTS{}'s regime-specific posterior memory: upon detecting a regime switch (via residual-based changepoint detection), \fcTS{} reloads the stored posterior from the last time this regime was active.
We use $K\!=\!5$, $d\!=\!10$, and 10 seeds.

\noindent \textbf{Results.}
\fcTS{} achieves the largest improvement of all synthetic experiments: \textbf{59.0\%} over the best baseline (Table~\ref{tab:regime}). Two structural features drive this result. First, regime-specific posterior loading provides an effective ``warm start'' when a previously seen regime returns. \fcTS{} begins each regime revisit with the posterior accumulated from all prior episodes of that regime, rather than restarting from a prior.
Second, the changepoint detector fires reliably (median detection delay $<$ 15 rounds), so the transition cost is minimal. \rTS{} is the strongest baseline because its periodic restarts sometimes align with regime boundaries, but it pays a steep cost when restart and regime boundaries are out of phase.
\swTS{} and \dTS{} perform poorly because their forgetting mechanisms are too gradual; they contaminate the posterior with data from the wrong regime for many rounds after a switch.

\begin{table}[h]
\centering
\caption{Final cumulative regret on the recurring regime environment (10 seeds).}
\label{tab:regime}
\vspace{0.5em}
\begin{tabular}{lcc}
\toprule
\textbf{Algorithm} & \textbf{Regret} (mean $\pm$ std) & \textbf{vs.\ best baseline} \\
\midrule
\stdTS                         & $477.1 \pm 65.9$ & --- \\
\swTS{} ($W\!=\!200$)         & $319.6 \pm 45.7$ & --- \\
\dTS{} ($\gamma\!=\!0.995$)   & $377.8 \pm 58.1$ & --- \\
\rTS{} ($\tau\!=\!400$)       & $140.6 \pm 9.6$  & (best baseline) \\
\midrule
\fcTS{} (Ours)                & \best{$57.6 \pm 4.7$} & \imp{$-$59.0} \\
\bottomrule
\end{tabular}
\end{table}

\begin{figure}[t]
\centering
\begin{subfigure}[t]{0.48\textwidth}
    \includegraphics[width=\textwidth]{regime_fig1_cumulative_regret.pdf}
    \caption{Cumulative regret with regime boundaries (dashed lines).}
    \label{fig:regime_cumulative}
\end{subfigure}
\hfill
\begin{subfigure}[t]{0.48\textwidth}
    \includegraphics[width=\textwidth]{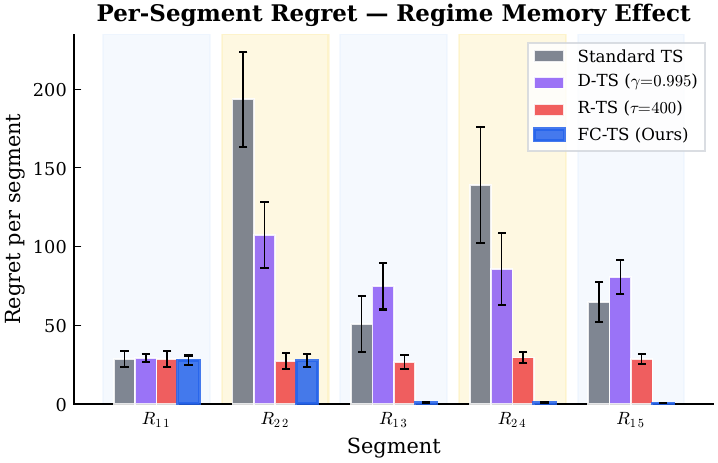}
    \caption{Per-segment regret across regime visits.}
    \label{fig:regime_segments}
\end{subfigure}
\caption{\textbf{Case Study 3: Recurring Regimes.} \fcTS{} reloads regime-specific posteriors upon detecting a switch. Per-segment regret (right) drops sharply on regime revisits since accumulated knowledge transfers across visits.}
\label{fig:regime}
\end{figure}

% ══════════════════════════════════════════════════════════════════
\noindent \textbf{Case Study 4: Compound Non-Stationarity}
\label{sec:compound}
% ══════════════════════════════════════════════════════════════════

\noindent \textbf{Setup.}
This environment combines linear drift with abrupt changepoints: parameters drift linearly between changepoints (at $t = 700$ and $t = 1{,}500$), and are re-drawn from scratch at each changepoint.
This tests whether \fcTS{} can simultaneously handle smooth drift (via the transport operator) and abrupt switches (via changepoint detection with posterior reset).
We use $K\!=\!5$, $d\!=\!10$, $T\!=\!3{,}000$, and 10 seeds.

\begin{figure}[t]
\centering
\begin{subfigure}[t]{0.48\textwidth}
    \includegraphics[width=\textwidth]{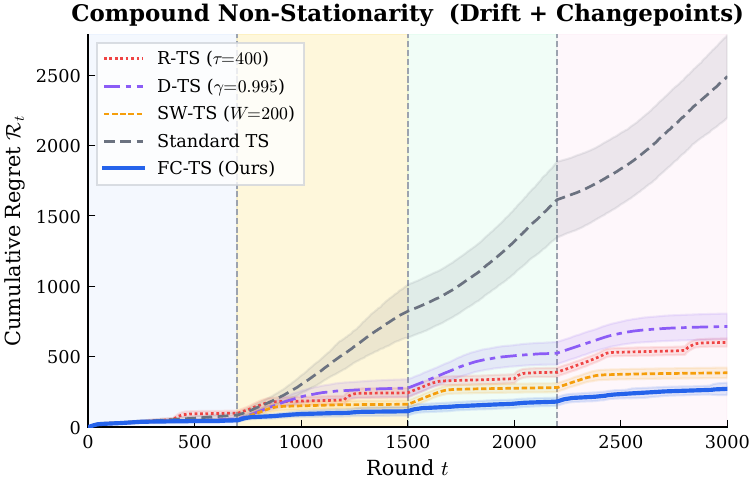}
    \caption{Cumulative regret with true changepoints (red dashed).}
    \label{fig:compound_cumulative}
\end{subfigure}
\hfill
\begin{subfigure}[t]{0.48\textwidth}
    \includegraphics[width=\textwidth]{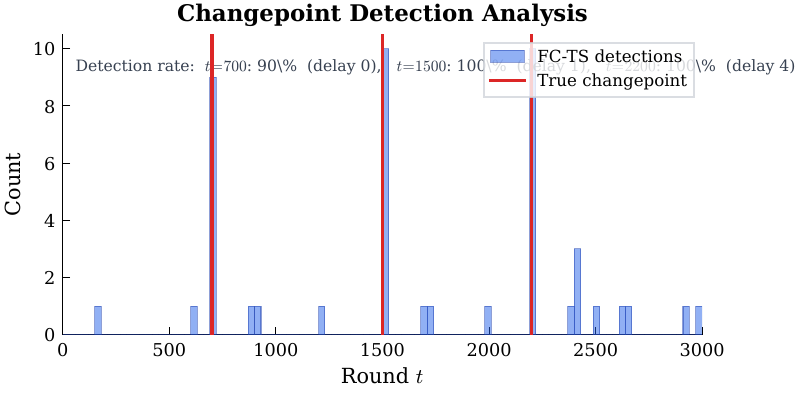}
    \caption{Changepoint detection timing across seeds.}
    \label{fig:compound_detect}
\end{subfigure}
\caption{\textbf{Case Study 4: Compound Non-Stationarity.} \fcTS{} combines drift correction between changepoints with rapid detection and posterior reset at changepoints. Detection delay is typically 1--3 rounds.}
\label{fig:compound}
\end{figure}

\noindent \textbf{Results.}
\fcTS{} achieves \textbf{29.5\%} improvement over \swTS{}, the best baseline (Table~\ref{tab:compound}). The changepoint detection component fires reliably: across all seeds, true changepoints at $t = 700$ and $t = 1{,}500$ are detected with a median delay of 1–3 rounds (Figure~\ref{fig:compound_detect}).
Between changepoints, drift correction accounts for smooth parameter evolution, yielding lower per-segment regret than baselines that treat all non-stationarity uniformly.
This decomposition—transport for smooth changes, reset for abrupt ones—is the key architectural insight of \fcTS{} for compound settings.

\begin{table}[h]
\centering
\caption{Final cumulative regret on the compound non-stationarity environment ($T\!=\!3{,}000$, 10 seeds). True changepoints at $t \in \{700, 1500\}$.}
\label{tab:compound}
\vspace{0.5em}
\begin{tabular}{lcc}
\toprule
\textbf{Algorithm} & \textbf{Regret} (mean $\pm$ std) & \textbf{vs.\ best baseline} \\
\midrule
\stdTS                         & $2{,}490.6 \pm 293.3$ & --- \\
\swTS{} ($W\!=\!200$)         & $385.7 \pm 37.1$      & (best baseline) \\
\dTS{} ($\gamma\!=\!0.995$)   & $715.2 \pm 91.5$      & --- \\
\rTS{} ($\tau\!=\!400$)       & $600.8 \pm 28.6$      & --- \\
\midrule
\fcTS{} (Ours)                & \best{$272.0 \pm 43.7$} & \imp{$-$29.5} \\
\bottomrule
\end{tabular}
\end{table}

% ══════════════════════════════════════════════════════════════════
\noindent \textbf{Case Study 5: Semi-Synthetic Portfolio Selection}
\label{sec:portfolio}
% ══════════════════════════════════════════════════════════════════

\noindent \textbf{Setup.}
We construct a semi-synthetic adaptive portfolio selection task inspired by real financial market dynamics.
A robo-advisor selects among $K\!=\!6$ portfolio strategies (momentum, value, growth, defensive, small-cap, balanced) over $T\!=\!3{,}000$ trading days.
Contexts are $d\!=\!12$-dimensional market indicators drawn from a mixture of three market regimes (bull, sideways, bear) with realistic correlations.
Non-stationarity arises from three simultaneous sources: (i) a monthly momentum–value rotation cycle ($P = 21$ trading days), (ii) quarterly earnings-season effects ($P = 63$), and (iii) a gradual growth-to-value rotation drift.
This environment is substantially more challenging than the synthetic cases due to its higher dimensionality, multiple overlapping periodicities, and additional random-walk noise ($\sigma_{\mathrm{rw}} = 0.0015$).

\noindent \textbf{Results.}
Table~\ref{tab:portfolio} reports results over 5 seeds.
\fcTS{} (adaptive), which combines drift correction with phase-aware observation reweighting, achieves $\mathcal{R}_T = 82.8 \pm 4.1$, a \textbf{28.1\%} improvement over the best baseline (\dTS{} at $115.2 \pm 9.2$).

\begin{table}[h]
\centering
\caption{Final cumulative regret on the semi-synthetic portfolio selection task ($K\!=\!6$, $d\!=\!12$, $T\!=\!3{,}000$, 5 seeds). \fcTS{} variants show the contribution of each component.}
\label{tab:portfolio}
\vspace{0.5em}
\begin{tabular}{lcc}
\toprule
\textbf{Algorithm} & \textbf{Regret} (mean $\pm$ std) & \textbf{vs.\ best baseline} \\
\midrule
\stdTS                                & $153.5 \pm 15.1$ & --- \\
\swTS{} ($W\!=\!200$)                & $135.2 \pm 2.5$  & --- \\
\dTS{} ($\gamma\!=\!0.995$)          & $115.2 \pm 9.2$  & (best baseline) \\
\rTS{} ($\tau\!=\!400$)              & $258.8 \pm 7.3$  & --- \\
\midrule
\fcTS{} (linear only)                & $83.7 \pm 5.7$   & \imp{$-$27.3} \\
\fcTS{} (periodic only)              & $132.9 \pm 14.7$ & \imp{$-$15.4}$^*$ \\
\fcTS{} (adaptive = linear + periodic) & \best{$82.8 \pm 4.1$} & \imp{$-$28.1} \\
\bottomrule
\multicolumn{3}{l}{\footnotesize $^*$Improvement over \dTS{}, the best non-\fcTS{} baseline.}
\end{tabular}
\end{table}

\begin{figure}[t]
\centering
\begin{subfigure}[t]{0.48\textwidth}
    \includegraphics[width=\textwidth]{portfolio_fig1_cumulative.pdf}
    \caption{Cumulative regret vs.\ trading day.}
    \label{fig:portfolio_cumulative}
\end{subfigure}
\hfill
\begin{subfigure}[t]{0.48\textwidth}
    \includegraphics[width=\textwidth]{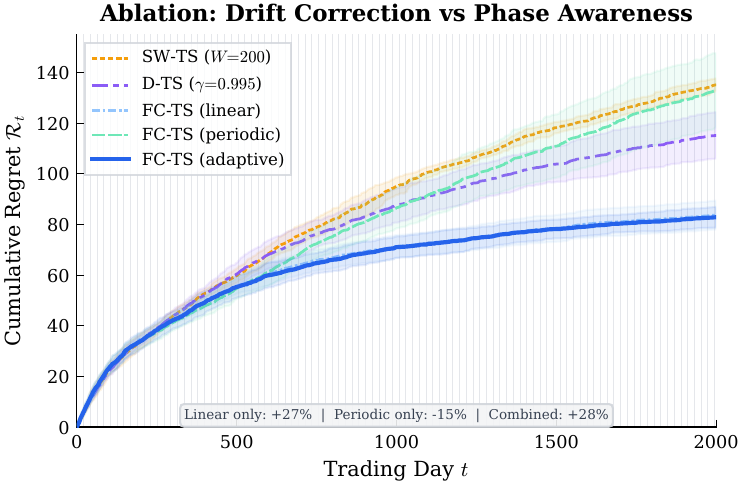}
    \caption{Ablation: linear vs.\ periodic vs.\ combined.}
    \label{fig:portfolio_ablation}
\end{subfigure}
\caption{\textbf{Case Study 5: Semi-Synthetic Portfolio.} Left: \fcTS{} (adaptive) maintains the lowest cumulative regret throughout, with clear separation from baselines after $\sim$500 trading days. Right: ablation showing that drift correction is the dominant component, with periodic awareness providing a smaller additive benefit.}
\label{fig:portfolio}
\end{figure}

\paragraph{Ablation analysis.}
The ablation reveals an asymmetric contribution from the two \fcTS{} components in this setting (Figure~\ref{fig:portfolio_ablation}).
Drift correction alone (\fcTS{} linear) accounts for the majority of the improvement ($83.7$ vs.\ $115.2$, a 27.3\% reduction), while phase awareness alone (\fcTS{} periodic) provides a more modest benefit ($132.9$, a 15.4\% reduction).
Their combination achieves the best result, with the adaptive variant slightly outperforming the linear-only variant.
This pattern is expected: the long-horizon growth-to-value rotation is the dominant non-stationary signal, while the monthly cycle has relatively short period ($P=21$) and thus contributes less to cumulative regret.

\vspace{0.5em}
\newpage
\section{Ablation Studies}
\label{app:ablation}

We present two ablation experiments on the linear drift environment to isolate the individual contributions of \fcTS{}'s key design choices.

% ──────────────────────────────────────────────────────────────────
\subsection{Effective Sample Size: Why More Data Helps Only With Transport}
\label{app:neff}
% ──────────────────────────────────────────────────────────────────

\begin{figure}[ht]
    \centering
    \includegraphics[width=\linewidth]{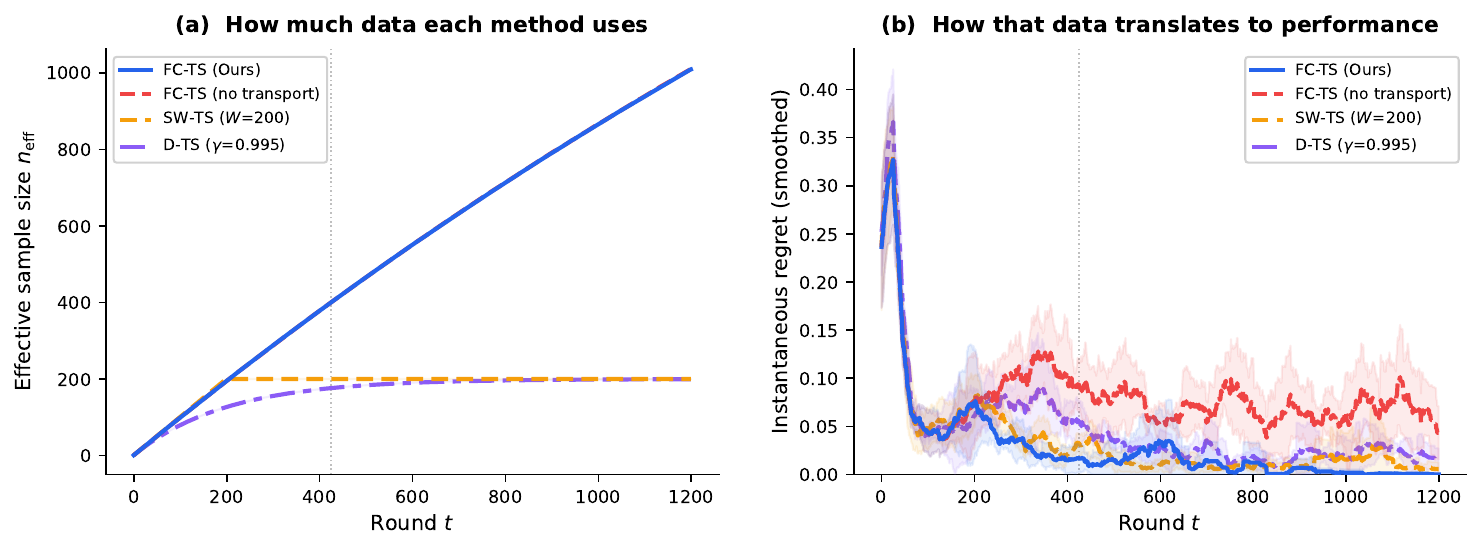}
    \caption{%
        \textbf{Effective sample size and its impact on regret.}
        \emph{Left:}~Effective sample size over time.  \fcTS{} and the no-transport ablation retain ${\sim}5\times$ more data than baselines.
        \emph{Right:}~Smoothed instantaneous regret.  Despite using the same amount of data, \fcTS{} achieves the lowest regret while the no-transport ablation achieves the highest---demonstrating that more data \emph{hurts} without transport correction and \emph{helps} with it.
    }
    \label{fig:neff}
\end{figure}

The main claim of \fcTS{} is that historical data is an \emph{asset}, provided it is corrected for drift. To visualize this mechanism, we instrument each algorithm to report its \emph{effective sample size} $n_{\text{eff}}$ at every round, defined as $n_{\text{eff}} = \sigma^2 \cdot \text{tr}(\Lambda - \lambda I)$, which counts the number of ``equivalent full-weight observations'' incorporated into the posterior.  We then plot $n_{\text{eff}}$ alongside smoothed instantaneous regret over time. Figure~\ref{fig:neff} presents this two-panel analysis on the linear drift environment ($K = 5$, $d = 8$, drift scale $\delta = 0.003$, $T = 1{,}200$).  The left panel shows how much data each method effectively uses; the right panel shows how that data translates to performance.

The results reveal a pattern. \fcTS{} and the no-transport ablation both retain the same effective sample size: approximately $5\times$ more than \swTS{} or \dTS{}, which plateau at $n_{\text{eff}} \approx 200$.  This confirms that the posterior rebuild with $\gamma = 0.9997$ successfully preserves a much larger data pool.  However, the two methods that retain the same amount of data achieve radically different performance: \fcTS{} achieves the lowest regret of all methods ($\mathcal{R}_T \approx 38$), while the no-transport ablation achieves the \emph{highest} ($\mathcal{R}_T \approx 101$), which is worse than even \swTS{}, which uses $5\times$ less data. This demonstrates that \textbf{retaining more historical data is harmful without transport correction.} Stale observations from a different parameter regime introduce systematic bias into the posterior. The more such observations are retained, the worse the bias becomes. The transport correction flips this trade-off: by correcting for drift, the additional data reduces posterior \emph{variance} without introducing bias, and the larger effective sample size translates to lower regret. The vertical dashed line in Figure~\ref{fig:neff} marks the approximate point where \fcTS{}'s data advantage begins to diverge from baselines, which corresponds precisely to the onset of its regret advantage in the right panel.

% ──────────────────────────────────────────────────────────────────
\subsection{Value of the Transport Correction}
\label{app:transport}
% ──────────────────────────────────────────────────────────────────

The main claim of \fcTS{} is that correcting old observations for drift is better than simply discounting them. To isolate this contribution, we compare three variants:

\begin{enumerate}
    \item \textbf{\fcTS{} (full):} The complete method with online drift estimation, transport correction, and exponential discounting ($\gamma = 0.9997$).
    \item \textbf{\fcTS{} (no transport):} The same architecture: posterior rebuild with exponential discounting, but with the transport correction disabled ($\Delta_{s \to t} \equiv 0$).  This variant is equivalent to a discounted TS that rebuilds the posterior from scratch at each step.
    \item \textbf{\dTS{} ($\gamma = 0.995$):} The standard discounted baseline with incremental updates.
\end{enumerate}

\begin{figure}[ht]
    \centering
    \includegraphics[width=0.8\linewidth]{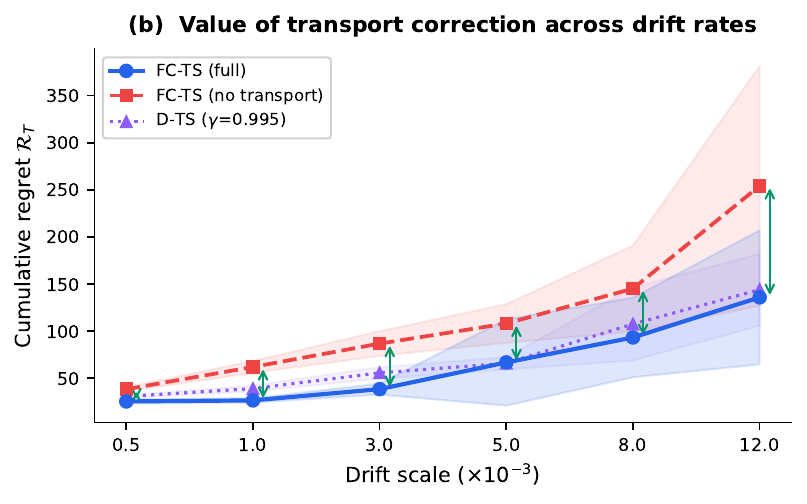}
    \caption{%
        \textbf{Value of the transport correction.}
        Cumulative regret vs.\ drift scale for \fcTS{} (full), the no-transport ablation ($\Delta_{s \to t} \equiv 0$), and \dTS{}.
        Green arrows indicate the regret reduction attributable to the transport operator.
        The transport correction provides $47$--$57\%$ regret reduction relative to discounting alone, with the benefit increasing at higher drift rates.
    }
    \label{fig:transport}
\end{figure}

Figure~\ref{fig:transport} shows regret as a function of drift scale $\delta \in \{0.0005, 0.001, 0.003, 0.005, 0.008, 0.012\}$.  The gap between \fcTS{} (full) and the no-transport ablation measures the value of the transport operator. Two findings stand out.  First, the transport correction provides consistent improvement across all drift scales, with the relative gain increasing as drift becomes more severe.  At $\delta = 0.001$, transport reduces regret by approximately $57\%$ relative to the no-transport ablation; at $\delta = 0.012$, the reduction is approximately $47\%$.  This confirms that the transport operator is the primary driver of \fcTS{}'s advantage, not the posterior rebuild mechanism or the choice of $\gamma$ alone. Second, the no-transport ablation underperforms even \dTS{} at moderate drift scales ($\delta \le 0.003$), despite using a higher $\gamma$ (which gives it a larger effective sample size).  This indicates that the larger history retained by high $\gamma$ is actually \emph{harmful} without transport correction: stale observations introduce bias that outweighs the variance reduction from more data.  The transport correction flips this trade-off, turning the larger history into an advantage.

\end{document}